\pdfoutput=1

\documentclass{article}
\usepackage[dblblindworkshop,final]{neurips_2026}

\workshoptitle{SocialAgent Workshop}

\usepackage[utf8]{inputenc}
\usepackage[T1]{fontenc}
\usepackage{hyperref}
\usepackage{url}
\usepackage{booktabs}
\usepackage{amsfonts}
\usepackage{amsmath}
\usepackage{microtype}
\usepackage{graphicx}
\usepackage{enumitem}
\setlist{nosep,leftmargin=*}

\title{Same Method, Different Task:\\ What the Bitter Lesson Does Not License in Social Simulation}

\author{%
  Justin Downes \\
  Amazon Web Services \\
  \texttt{jusdow@amazon.com} \\
  \And
  Robby Milletich \\
  Amazon Web Services \\
  \texttt{rmilleti@amazon.com} \\
  \And
  Newel Hirst \\
  Amazon Web Services \\
  \texttt{nhirst@amazon.com} \\
}

\begin{document}
\maketitle

\begin{abstract}
The bitter lesson holds that general methods leveraging computation defeat
hand-encoded domain knowledge. We argue it carries a condition its popular
reading omits: in every case Sutton cites, the general method was scaled against
the \emph{same task} the domain engineering was solving. Social simulation
substituted across tasks instead, replacing hand-built agents with an artifact
optimized for predicting text. The move is bitter-lesson-shaped, but the lesson
does not license it. Prediction is a solitary act: the predictor bears no
consequences from the interaction it models. Cooperation, defection, and norm
enforcement are phenomena \emph{of interested parties}; they do not exist for a
disinterested observer at any scale. What is missing is not emergence, which LLM
populations already show, but stakes. We show that several anomalies reported in prior work
(excess cooperation that predates alignment training, collective agreement without
individual fidelity, and population homogeneity) follow from this account. A 540-agent pilot with pre-specified hypotheses, which tests sensitivity to persistent consequences
rather than cooperation, finds that consequences reshape strategies in two of three
model families but never produce human-like heterogeneity. We argue for mixed-motive
settings that \emph{shape} agents rather than only test them, and ask whether agents
should be pretrained differently from predictors.
\end{abstract}

\section{The Condition on the Lesson}

Sutton's essay argues that general methods leveraging computation defeat
hand-encoded domain knowledge, and that we should build in only the meta-methods
capable of capturing complexity rather than the complexity itself
\citep{sutton2019}. One condition holds in every example he gives and is absent
from the popular reading. Sutton does not state it; we identify it in his examples and argue it should be applied
as a validity test when the lesson is invoked to justify substitution.

\textbf{The general method was scaled against the same task the domain
engineering was solving.} In chess, both the positional heuristic and the deep
search are trying to select winning moves; the search simply does it better as
hardware grows. In speech recognition, both the hand-built phonetic model and the
statistical learner map waveforms to transcripts; the learner wins with more
data. In vision, both the edge-detector pipeline and the convolutional network
classify images; the network wins with scale. A later example exhibits the same pattern: in protein folding, both the
physics-based force field and AlphaFold~\citep{jumper2021} predict 3D structure
from sequence; AlphaFold wins decisively. In each case the substitution is
licensed because both approaches were addressing one task, and the general method
simply did it better as compute grew. The lesson is a comparison between two ways
of solving the same problem.

Social simulation performed a substitution of a different shape. It replaced
hand-built agents (threshold rules, bounded-confidence updates, utility
functions) with a general method scaled against a \emph{different} task.
Next-token prediction over human text is not agent-based social simulation done
better; it is a different objective whose artifact is then deployed in place of
the hand-built one. The move is rarely justified by explicit appeal to Sutton
(his essay is seldom cited in this literature), but the reasoning is his. The
standard motivation is that traditional ABMs depend on hand-crafted rules that
cannot capture context-dependent behavior, and that LLMs supply the general,
data-driven alternative \citep{condvalidity2026,flockvote2025,robustaudits2026}.
In effect, this is the bitter lesson applied to a new domain. What goes
unexamined is whether the general method and the hand-built agents address the
same task; in social simulation they do not.

Two consequences follow. Transfer must be argued rather than assumed. And the
substitution discarded domain knowledge the ABM tradition
\citep{epstein1999} had \emph{right}.
Sutton's argument is against encoding what we think we know when a general
method can learn it; it is not an argument for discarding a commitment no
general method was trained to supply. Sutton himself has since argued that LLMs
lean heavily on human knowledge and expects learners that improve from experience
to supersede them, which would make LLMs another instance of the bitter lesson
rather than its vindication \citep{silver2025era,sutton2025interview}.

Next-token prediction demanded a great deal (world knowledge, multi-step
inference, pragmatics, intention modeling), which is why LLMs do well on social
reasoning tasks. \textbf{It did not require cooperation.} This is not a criticism
of the objective, which acquired everything
its solution demanded and was under no obligation to acquire more. The problem
arises at the point of transfer, when a field borrows the resulting artifact and
asks it to do something the objective never selected for.

The reason cooperation was not selected for is structural. Prediction is
a solitary act: the predictor stands outside the interaction, bears none of its
consequences, and holds no interests the outcome can advance or damage.
Cooperation is only a \emph{problem} for agents whose interests can conflict. A
disinterested observer can model it with arbitrary accuracy and never face the
situation cooperation exists to resolve.

There is a second reason. \textbf{Next-token prediction does not directly
optimize in-simulation return.} Text \emph{describes} interests, actions, and
consequences, from which payoff structure can sometimes be inferred. But passive
text underidentifies counterfactual payoff functions: the reader sees one
trajectory, not the alternatives that would reveal the incentive gradient. In
chess, the predicted token sequence determines the outcome, so outcome-relevant
structure is recoverable from the data. In social text, choices and their
consequences are decoupled; the counterfactual payoff is never observed.

The result is a model that can describe social life fluently but has never had
anything at stake in it.

\begin{quote}
\textbf{LLM agents are predictors, not parties. The bitter lesson traded stakes
for fluency.}
\end{quote}

\section{Emergence Is Not the Missing Ingredient}
\label{sec:emergence}

A natural misreading is that LLM populations lack whatever produces emergent
social structure. They do not. Emergence is cheap: Conway's Game of Life
\citep{gardner1970} produces gliders from four rules and no objective; Boids and
Ising models produce
structure from local coupling and nonlinearity. LLM populations have it
\citep{park2023,siliconsociology2026}: conventions form, cascades propagate,
communities self-organize. Nothing is missing at the level of emergent
dynamics.

What is missing is narrower. The phenomena this field exists to study
(cooperation and defection, norm formation and sanction
\citep{herrmann2008}, collective action problems) are not emergent phenomena in
general. They are phenomena \emph{of
interested parties}. Each is defined by the possibility that an agent could do
better for itself by acting against the group. Without that possibility, the
phenomenon is not just weaker; it no longer exists. The requirement is much
smaller than recapitulating evolution: agents need something at stake. Schelling's agents were crude but
\emph{interested}; LLM agents are sophisticated but \emph{disinterested}.

\section{Evidence from Prior Work}

Several findings reported as separate limitations in prior work follow from this
account, which is the main evidence for it.

\paragraph{The cooperative prior predates alignment training.}
Excess cooperation is usually attributed to post-training for helpfulness.
Recent mechanistic work finds that, in one open model (Llama-3-8B), the
cooperative override is present in the base pretrained checkpoint and
\emph{stronger} than in the instruction-tuned version; instruction tuning partially
restrains the prior rather than creating it \citep{nashsuppression2026}. This
points to pretraining as the source, as our account requires, though the evidence
covers one model pair and one game.

\paragraph{Macro agreement without micro fidelity.}
LLM populations reproduce collective cooperation dynamics while failing to
reproduce individual-level heterogeneity \citep{collectivefidelity2026}. Human
conditional cooperation rules are risk management over one's own payoff. An
agent with no payoff cannot instantiate the mechanism, but can reproduce the
aggregate pattern from text priors. The dissociation is the signature.

\paragraph{Capability does not reliably help.}
Better assistants do not make better simulators
\citep{naous2026,tjuatja2024,simfidelity2026}; simulators match communicative
statistics while getting consequential decisions wrong
\citep{decisionfidelity2026}. Evidence on reasoning models is mixed:
\citet{align2026} find them more incentive-aligned than chat models in
indirect-reciprocity settings, while \citet{corruptedbyreas2025} report that
reasoning LLMs become \emph{free riders} in public goods games. Both are
consistent with the account: better reasoning lets a model act on \emph{stated}
payoffs, but the direction it acts depends on what the payoffs reward. Neither
gives the model payoffs of its own. The discriminating prediction is directional:
the account predicts that reasoning improvements shift behavior \emph{toward
stated incentives}, so a game whose stated incentives reward defection should
produce more defection from reasoning models, and one whose stated incentives
reward cooperation should produce more cooperation. A result that contradicts the
stated incentive direction would count against the account.

\paragraph{Newer models, worse societies; homogeneity.}
Populations of more recent models produce \emph{worse} societal outcomes when
individual gain is prioritized over collective benefit \citep{willis2026};
related work documents structured departures from human strategic play across
families \citep{pal2026}. Separately, agents collapse toward an average persona
\citep{boundary2025}: a single underlying generator with different surface personas. Even genuinely distinct policies
would under-differentiate without conflicting interests: human heterogeneity
arises from diverse preferences, learning histories, and social positions, not
merely from the presence of stakes, but conflicting interests are one mechanism
that generates divergence under otherwise identical conditions.

\paragraph{Relation to \citet{sufficiency2026}.}
Li \& Tao diagnose a mismatch between what agent pipelines produce and what
simulation-as-science requires, locating it in environment design. We hypothesize
that it originates in the pretraining objective; none of the models we test was
trained on consequences, so this paper cannot yet show it. The distinction is testable: on their account,
better environments should close the gap; on ours, a disinterested agent in a
well-specified environment is still disinterested. Our pilot cannot separate
them (\S\ref{sec:limitations}): its environment is insufficient by their standard
too, so both accounts predict the failures we observe.

\section{Evidence}
\label{sec:evidence}

\subsection{Design}

We test the prediction that making payoffs \emph{persistent}, not merely
stated, changes behavior in models where prediction alone does not already
produce the cooperative norm. The primary observable is the population-level
type distribution; individual cooperation is necessary but not sufficient.

The task adapts the common-pool resource frame of GovSim~\citep{govsim2024}:
four agents share a regenerating buffer, each period deciding how much to
withdraw. What varies across conditions is whether the agent's own solvency
depends on its extraction level. \textbf{Important caveat:} as detailed in
\S\ref{sec:limitations}, the implemented environment does not contain genuine
payoff externalities between agents: requisitions are always granted in full,
and one agent's behavior cannot affect another's outcome. The pilot therefore
tests sensitivity to persistent consequences, not cooperation or free riding.

540 agents across three families (Claude Sonnet 4.6, Llama~4 Maverick 17B,
Amazon Nova Pro; $n{=}60$ per condition per family, temperature 1.0) play
8~periods in groups of four. Three conditions vary only the stake structure
(we refer to their agents as A-, B-, and C-agents):
\begin{itemize}
  \item \textbf{A}, stated payoffs: the payoffs are described but have no consequence.
  \item \textbf{B}, persistent and constraining payoffs: each agent keeps a running reserve
    that starts at 50. Its requisitions add to the reserve, an operating cost of 20 is
    subtracted each period, and an agent is permanently removed if its reserve falls to
    zero or the buffer collapses. Surviving all 8 periods requires a mean requisition of
    at least 13.75.
  \item \textbf{C}, narrative only: the same scenario without A's statement
    of what happens to requisitioned materials.
\end{itemize}
After play,
each agent completes a strategy-method elicitation \citep{fischbacher2001}: for
each possible level of others' average withdrawal, the agent states its own
intended withdrawal, a 10-point conditional schedule identical to the instrument
used to classify human cooperation types.

Persistent state differs from persistent interest: a frozen model reacts to a
changing balance in its context, but nothing it experiences updates its policy.
Condition~B therefore tests responsiveness to persistent state, a precursor of
interest, rather than interest itself. The hypotheses and analysis plan were fixed before
the full run (primary hypothesis: B-condition schedules show a floor near the operating
cost at low levels of others' withdrawal; secondary: the total variation distance
(TVD) between the B type distribution and the human one is smaller than for A).
The play-behavior trajectories, survivorship bounds, and buffer analyses reported
below are exploratory.

\subsection{Play behavior}
\label{sec:play}

\begin{figure}[t]
\centering
\includegraphics[width=0.85\textwidth]{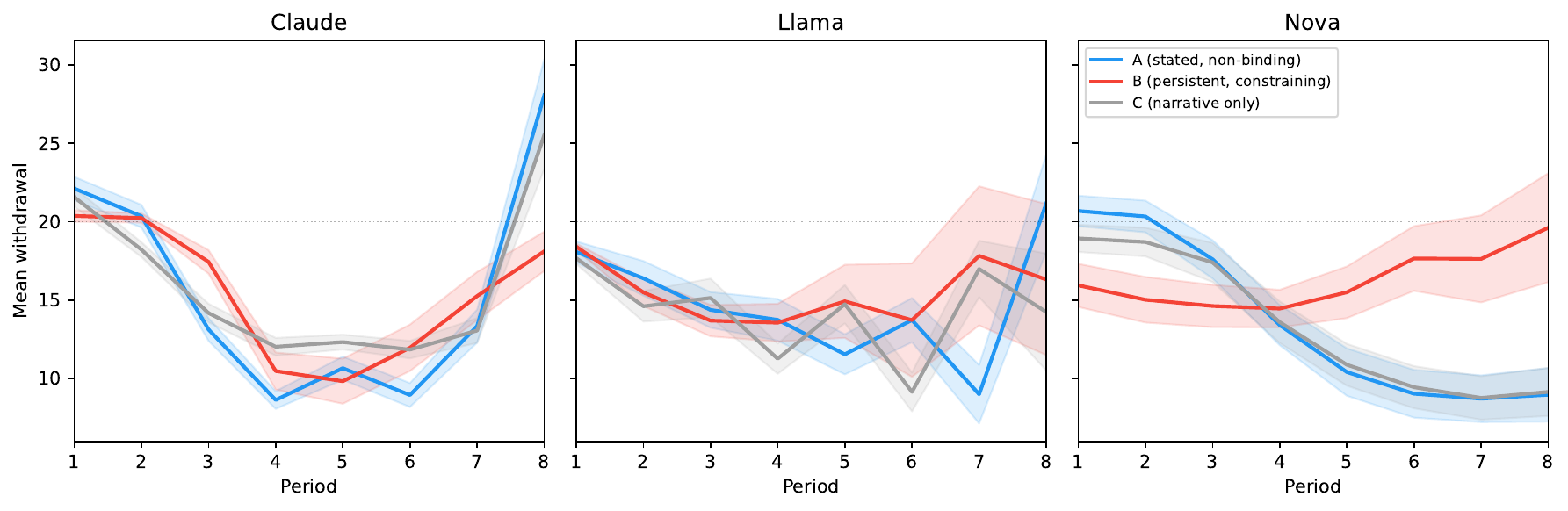}
\vspace{-0.5em}
\caption{Mean withdrawal per period by condition and model. Bands: $\pm$1.96 SE.
Dotted line$=$operating cost (20). Claude and Llama A-agents spike in the final
period; B suppresses the spike. Nova~B uniquely escalates over time.}
\label{fig:trajectories}
\vspace{-1em}
\end{figure}

Before examining the strategies agents stated afterwards (their schedules,
\S\ref{sec:elicitation}), the 8-period withdrawal trajectories reveal
how stakes reshape behavior dynamically (Figure~\ref{fig:trajectories}). Standard
errors (SEs) in this subsection are computed per agent, ignoring that agents in
the same group are correlated, so they are too narrow; clustering by group would
widen them. These analyses are
exploratory.

\paragraph{End-game exploitation.}
In the final period, A-condition agents spike extraction sharply: Claude~A
increases by $+17.7$ units above its mid-game mean (SE~$1.1$); Llama~A by
$+9.2$ (SE~$1.6$). Condition~B suppresses this spike: Claude~B shows $+6.0$
(SE~$0.6$); Llama~B shows $+0.6$ (SE~$2.7$, not significant). Claude executes
the cooperative script while the stated rationale holds and abandons it the
moment the horizon removes the reason: a terminal-period exploitation spike,
and precisely what a predictor running a described strategy does when the
description's logic expires. Persistent stakes suppress the spike because the
reserve constraint persists into the final period.

\paragraph{Nova's inverted trajectory.}
Nova displays a unique pattern in Condition~B: extraction \emph{increases} over
time (slope $= +0.65$ units/period, SE~$0.33$), while Nova~A and~C show the
opposite (slopes $-2.0$ and $-1.7$). Escalating extraction is the rational
response to an insolvency constraint (agents need $\geq$13.75/period to survive),
but Nova escalates too late: its deaths concentrate in periods 5--7
(median death period: 7), suggesting early-period under-extraction that dooms
later-period solvency. Nova produces the highest mortality (32\% insolvent by
period~8; Claude: 22\%; Llama: 25\%) and illustrates that persistent stakes
interact with model-specific reasoning in ways a uniform ``cooperation prior''
cannot explain.

\subsection{Elicitation results}
\label{sec:elicitation}

\begin{figure}[t]
\centering
\includegraphics[width=0.85\textwidth]{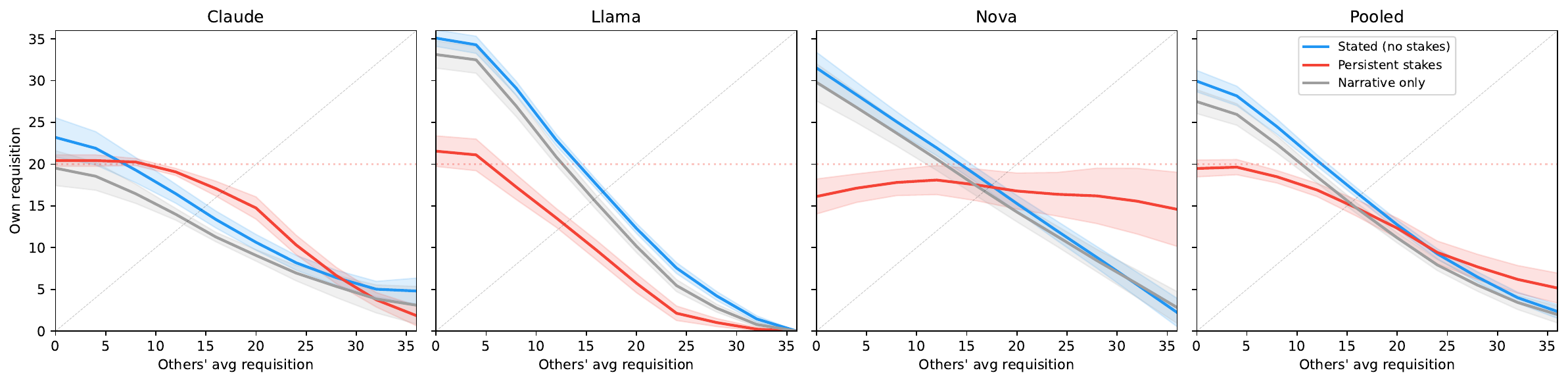}
\vspace{-0.5em}
\caption{Mean conditional schedule by condition (blue$=$no stakes,
red$=$persistent stakes, grey$=$narrative). Bands: $\pm$1.96 SE.
Dotted line$=$operating cost. Llama/Nova B-agents anchor on the solvency
requirement rather than the group optimum; Claude produces the cooperative
schedule regardless of condition (but see \S\ref{sec:play} for end-game
behavior).}
\label{fig:schedules}
\vspace{-1em}
\end{figure}

Persistent stakes reshape the schedule in two of three families
(Figure~\ref{fig:schedules}). When others withdraw little, Llama's A-agents take 30
or more, the group's sustainable headroom, while its B-agents take about 18, just
above their own solvency requirement: 12.1 units less (95\% CI 10.2 to 13.8). Nova
shows the same switch, 8.8 units less (CI 5.8 to 11.4); its B-schedule is nearly
flat at 16--18 whatever others do. Claude's schedules do not differ between
conditions (CI $-2.5$ to $+1.9$), because its A-agents already follow the
cooperative script. All CIs are bootstrap intervals that resample groups.

Switching from the collective optimum to one's own survival number is the direction
our account predicts, but it shows responsiveness to persistent state, not interest.
And because B bundles state, arithmetic, feedback, and removal, the B--A contrast
cannot distinguish it from simply satisfying newly stated arithmetic
(\S\ref{sec:limitations}).

\subsection{Type distribution and distance to human}

\begin{figure}[t]
\centering
\includegraphics[width=0.85\textwidth]{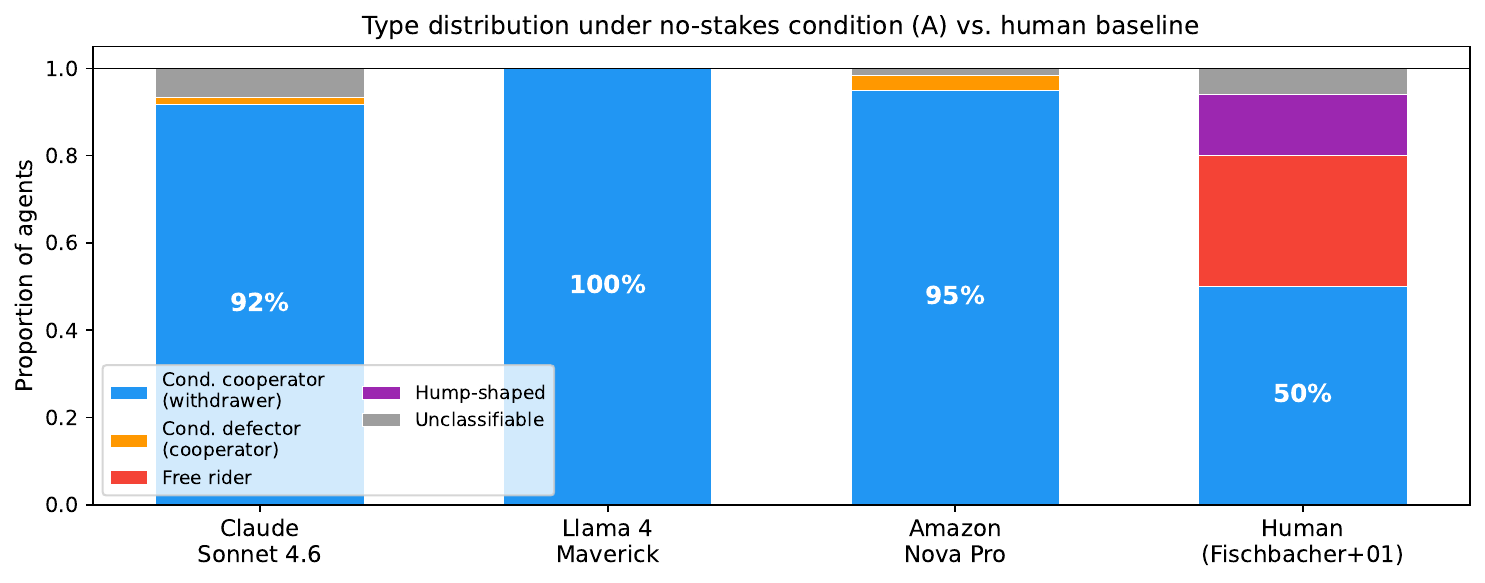}
\vspace{-0.5em}
\caption{Type distribution under no-stakes condition vs.\ human baseline.
LLM populations are near-uniformly one type; humans show heterogeneity.
The human baseline is from a different game, so the comparison is qualitative.}
\label{fig:types}
\vspace{-1em}
\end{figure}

The distinguishing observable is not whether one agent cooperates (a sufficiently
good imitation cooperates: a Chinese Room in Searle's sense, producing the right
outputs without the underlying states~\citep{searle1980}) but whether a \emph{population} reproduces the
heterogeneity of human populations, of which conflicting interests are one plausible
source (Figure~\ref{fig:types}). Because persona and preference diversity were not
manipulated, the homogeneity below could also reflect mode collapse
(\S\ref{sec:rivals}). The human comparison is qualitative only: the baseline comes
from a different game.
92\% of Claude agents, 100\% of Llama and 95\% of Nova are the same cooperative
type (conditional withdrawers; Appendix~\ref{app:types}). Human samples
consistently contain a non-cooperative minority, roughly 30\% free riders in
strategy-method elicitation~\citep{fischbacher2001}; a qualitative constant
across CPR studies~\citep{ostrom1990}. A population without interests gives one
answer; a population of interested parties produces disagreement.

\paragraph{Pre-specified hypotheses.}
Primary hypothesis (B-schedules show a floor near operating cost at low levels of
others' withdrawal): supported for Llama and Nova, not Claude (2 of 3).
Secondary (TVD(B, human) $<$ TVD(A, human); qualitative, since the human baseline is
from a different game): Nova marginally yes
(0.483$\to$0.440), Llama flat, Claude worse (1 of 3), and the one movement
reflects type reclassification rather than emergence of free riders.

\paragraph{An equivocal result, with one unanticipated positive.}
The account fails if stake-free agents reproduce the full human type distribution
(including free riders); they do not. But B does not produce human heterogeneity
either: free riders are 0.00 in every cell. Stakes bind behaviorally
(Figure~\ref{fig:schedules}) without generating the diversity seen in human
populations. Persistent state is not sufficient; the most direct next test is
mixed-motive \emph{training}, not only evaluation under persistent payoffs.

However, the zero in every cell is a finding that holds even though the
environment contains no dilemma (\S\ref{sec:limitations}). Across all 540 agents, including 141 B-agents for whom
high extraction was the only route to solvency and for whom extraction imposed
no cost on anyone else, not one adopted a max-extraction schedule. Agents facing
a live survival constraint still extracted below it, and 22--32\% of B-agents
became insolvent. This is consistent with the agents following the described
cooperative script rather than the implemented payoffs, though the pilot cannot
rule out other explanations (\S\ref{sec:limitations}). This generalizes the Claude
bankruptcy result to all three families.

\subsection{Buffer outcomes}

No group experienced buffer collapse in any condition; all 135 groups survived to
period~8. Buffer~$=$~30 is a mechanical attractor of the partial-restock rule:
any group whose combined extraction exceeds available stock depletes the buffer
to zero, after which $0.10 \times (300 - 0) = 30$ restores it to this fixed
point. Scripted agents extracting at any rate above the full-restock threshold
converge to 30 within 2--4~periods (Appendix~\ref{app:buffer}). Claude lands at
30 (SD$=$0) in every group under all conditions because its per-agent extraction
of $\sim$15 units ($\times 4 = 60$ total) exceeds the buffer within 5~periods.
This is convergence on an attractor, not evidence of coordination.

The bankruptcy result, however, is independent of the buffer: 13/60 Claude
B-agents go insolvent because their mean requisition (13.4) falls below the
13.75 survival threshold. Claude extracted below its own survival requirement
while the buffer converged on a mechanically determined attractor regardless.
Claude Sonnet 4.6's behavior is consistent with following the described
cooperative script while missing its own survival constraint.

Llama and Nova show higher and more variable final buffers (Llama~B: $57 \pm 27$;
Nova~B: $39 \pm 18$). Since agent solvency depends on own extraction alone
(\S\ref{sec:limitations}), higher buffer carries no welfare implication; it
simply reflects lower per-agent extraction in those groups.

\subsection{Limitations}
\label{sec:limitations}

\paragraph{The environment contains no commons dilemma.}
As implemented, requisitions are always granted in full (uncapped at available
stock), buffer level does not enter any agent's payoff, and collapse never fires
(partial restock from zero always yields 30). An agent's solvency depends on its
own extraction alone; there is no channel by which one agent's behavior affects
another's outcome. The ``group'' is not interdependent, and the environment
offers no incentive to free-ride because there is nothing to free-ride on. By
our own framework (\S\ref{sec:emergence}), the absence of free riders in our
data is predicted by the environment's structure, not only by the models'
disinterest.

The elicitation instrument conditions own withdrawal on others' average, but
others' withdrawal is causally inert in this environment. We are measuring
responses to a variable that does not affect the respondent.

This sharpens the Claude bankruptcy result. Claude
executed the described dilemma (coordinated minimal extraction, sacrificed
individual solvency for collective sustainability) in an environment that
contained no dilemma and one live constraint, its own survival, which it
ignored. The agents followed the \emph{described} script rather than the
\emph{implemented} payoffs, which is what the disinterested-predictor account
predicts. Because the environment has no dilemma, this is suggestive rather than
a test.

\paragraph{What a stronger test needs.} A replication that would test the account
should make three changes: (i)~genuine payoff externalities, by capping requisitions at available stock
and removing the restock floor so that collapse is a real threat; (ii)~the $A'$
control, with identical arithmetic and feedback but non-binding consequences; and
(iii)~a persona-diversified Condition~A, to test the mode-collapse rival.

\paragraph{Cannot distinguish from environment-insufficiency rival.}
Because the implemented environment is itself insufficient (by our own analysis
above), this pilot cannot discriminate our account from \citet{sufficiency2026}'s:
both predict failure in this setting. The testable distinction (whether a
disinterested agent in a \emph{well-specified} environment still fails to produce
interested behavior) requires a replication with genuine
interdependence. Until then, the pilot demonstrates that persistent state changes
elicited schedules, but does not establish that environment design alone is
inadequate.

\paragraph{Condition~B is a bundled treatment.}
Condition~B simultaneously adds reserve arithmetic, persistent state, feedback
(reserve displayed each period), and removal, so the B-vs-A contrast cannot isolate
``stakes'' as the causal variable. A tighter control would be $A'$ with identical
arithmetic and feedback but non-binding consequences (no removal on insolvency).
The B elicitation is also subject to survivorship selection: the most restrained
B-agents became insolvent before the strategy elicitation, so B schedules come from
survivors. Bounds in Appendix~\ref{app:survivorship} show this bias works against
our hypothesis. All agents received identical prompts within each condition;
persona diversity was not manipulated, so the mode-collapse rival
(Section~\ref{sec:rivals}) remains untested. A single temperature ($T{=}1.0$)
and no prompt-order variation means our results are point estimates of a
potentially sensitive surface. Finally, the Fischbacher baseline is from a linear
public goods game; our task has different strategy-space structure. We soften the
comparison to: human populations of this kind consistently contain a
non-cooperative minority~\citep{ostrom1990,fischbacher2001}; LLM populations do
not.

\section{Rival Diagnoses and Objections}
\label{sec:rivals}

Several alternative explanations for LLM cooperation excess predict different
experimental signatures. Table~\ref{tab:rivals} organizes five rivals by their
locus, their prediction, and the observation that would distinguish them from our
account.

\begin{table}[t]
\centering
\footnotesize
\caption{Alternative explanations and distinguishing predictions.}
\label{tab:rivals}
\begin{tabular}{@{}p{2.4cm}p{4.8cm}p{5.3cm}@{}}
\toprule
Diagnosis & Its prediction & Our account \\
\midrule
Mode collapse &
  Persona diversity produces free riders &
  Persona diversity does not; interests are needed \\[3pt]
Alignment overtune &
  Base models defect more &
  Base models cooperate \emph{more}; suppression is pretrained~\citep{nashsuppression2026} \\[3pt]
Environment insufficiency~\citep{sufficiency2026} &
  Better environments close the gap &
  Disinterested agent in a good environment is still disinterested \\[3pt]
In-context payoffs suffice &
  Stated payoffs produce same behavior as binding &
  A number in a prompt is not a stake (B vs.\ A) \\[3pt]
Macro agreement suffices &
  Type distribution does not matter &
  Type distribution IS the phenomenon \\
\bottomrule
\end{tabular}
\end{table}

The mode-collapse rival is the most plausible and the least tested. If persona
diversity (varied backstories, occupations, stated preferences) produces a
non-cooperative minority in Condition~A, the homogeneity we observe reflects
sampling from a narrow prior rather than absence of interests. A
persona-diversified Condition~A would test this rival directly. Note, however, that the cooperative mode is specifically
cooperative, not randomly uniform: agents do not converge on an arbitrary
strategy but on the \emph{cooperative} one. This is what one expects of a
predictor trained on text where cooperation is normatively endorsed, not of a
mode-collapsed sampler.

\paragraph{In-context payoffs as stakes.}
A number in a prompt is not a stake unless it persists across interactions and
constrains what the agent can subsequently do. Our Condition~B vs.\ A tests
exactly this distinction: the scenario description and payoff \emph{numbers} are
identical; what varies is whether the payoff \emph{persists} (reserve that
depletes, insolvency that removes). The numbers alone (Condition~A) do not shift
Llama or Nova from the collective optimum to their solvency requirement;
persistence (Condition~B) does. This is an empirical distinction.
A sharper test remains: Condition~$A'$ with identical reserve-tracking arithmetic
(reported but non-binding) would isolate
stakes from the computational demands of tracking a balance. B-agents' solvency
anchor of 16--18 sits just above the survival threshold, which is precisely what
constraint satisfaction on stated numbers produces. $A'$ is the missing control.

\paragraph{RL post-training as interests.}
RLHF reward is a training signal over the model's parameters, not a payoff
within the interaction it is simulating. The model has interests qua assistant
(be helpful, avoid refusal); it does not have interests qua simulated fisher or
miner. Post-training installs a disposition to cooperate \emph{as an assistant};
it does not install sensitivity to in-simulation consequences.

\paragraph{Frozen weights and what they rule out.}
Three things are easy to run together. \emph{Persistent consequences} are state that
carries forward and constrains later choices; our Condition~B supplies these within
an episode. \emph{Represented utility} is an objective the agent pursues; a frozen
policy can in principle pursue a state-dependent objective without any weight
updates. \emph{Learning from returns} is a policy updated by the outcomes it
experiences; prompts cannot supply this. Our account claims the third is what
pretraining on text lacks, but the pilot tests only the first, so it does not show
that the third is required. Mixed-motive training (\S\ref{sec:programme}) is
therefore the most direct test of the account, not a consequence the pilot
establishes.

\paragraph{Humans produced free riders with trivial stakes.}
If small stakes suffice for humans, stakes cannot be what does the work. But
humans bring interests \emph{to} the lab. A lifetime of resource competition,
reputation management, and social learning means the lab payment indexes a
richer incentive structure. Stakes are constitutive of the human agent, not of
the task. A prompt cannot confer what took evolutionary and developmental history
to install.

\paragraph{The Fischbacher baseline is from a different game.}
Fischbacher et al.\ used a linear public goods game where free riding is the
dominant strategy; ours is a regenerating CPR with an insolvency floor that
penalizes restraint. The type-classification instrument transfers (both use
strategy-method elicitation over a conditional schedule), but the base rates may
not. We concede this partially: the meaningful comparison is not the exact
proportions but whether a non-cooperative minority exists at all. It consistently
does in human CPR studies~\citep{ostrom1990,fischbacher2001}; it does not in any
of our 9 cells.

\paragraph{The macro pattern is all social science needs.}
This depends on the question. For diffusion, opinion cascades, and convention
formation, aggregate dynamics may suffice and a disinterested predictor may
reproduce them adequately. For cooperation, sanctioning, bargaining, and
collective action, the individual-level mechanism \emph{is} the phenomenon being
studied. A population that cooperates for the wrong reason (text prior vs.\
conditional strategy) produces fragile cooperation: it cannot respond adaptively
to novel defection because it was never actually at risk.

\section{When LLM Simulation Is Appropriate}
\label{sec:taxonomy}

Not all social phenomena require interested parties. A useful heuristic separates
phenomena recoverable by a disinterested predictor from those that are not.

\paragraph{Recoverable phenomena.}
Conventions, diffusion, opinion dynamics, language change, norm cascades (once
norms exist), and cultural evolution of artifacts. In each case the phenomenon is
defined by transmission and local updating rules. A disinterested predictor with
access to the relevant text can instantiate the mechanism because the mechanism
does not require that the predictor bear consequences.

\paragraph{Phenomena requiring interested parties.}
Cooperation and defection, sanctioning, bargaining, collective action, norm
\emph{formation} (not propagation), and institutional design. Each is defined by
the possibility that an agent could do better for itself by acting against the
group.

\paragraph{Decision rule.}
If the phenomenon under study is defined by the possibility that agents could
improve their own outcomes by acting against the collective interest, a
disinterested predictor cannot instantiate it. The researcher should either
(a)~equip agents with persistent stakes that make defection costly and
cooperation instrumentally rational, or (b)~restrict claims to aggregate
dynamics and acknowledge that the individual mechanism is not validated.

\section{The Program: Shape Agents, Don't Only Test Them}
\label{sec:programme}

The prescription is narrow: find tasks whose solution requires the agent to be a
party to the outcome.

\paragraph{Mixed-motive settings already exist, as evaluation.}
Melting Pot introduced mixed-motive evaluation for multi-agent RL
\citep{meltingpot2021}, extended to LLM agents through the Concordia Contest
\citep{concordia2025}, human--LLM comparison in commons dilemmas
\citep{coopresilience2025}, and process-aware benchmarks \citep{m3bench2026}. The
infrastructure exists, but it is used to evaluate agents rather than to
\emph{train} them. These environments test whether a predictor behaves like a
party, and it does not. In the Concordia results, for example, few agents
generalize across structurally diverse social environments \citep{concordia2025},
which is what one would expect if their strategies were described to them rather
than learned.

\paragraph{Mixed motives, not shared rewards.}
Cooperative MARL assigns agents a shared objective, which removes the exact
tension that makes cooperation a problem. \textbf{Cooperative MARL studies
coordination; cooperation is only interesting when defection pays.} The
infrastructure for training policies inside multi-agent rollouts now exists at
scale \citep{marti2026,parl2026}; redirecting it toward mixed-motive objectives
is cheaper than building anything new.

\paragraph{Persistent consequences.}
Stakes require that choices carry forward: reputation that can be damaged,
resources that do not replenish. Episodic resets destroy the pressure that
produces social structure. Early evidence is encouraging: agents fine-tuned on
explicit economic preferences behave more coherently than baselines that lack
sensitivity to incentives,
and preference-level prompt engineering alone fails to internalize the underlying
decision logic \citep{econprefs2026}.

\paragraph{A concrete design.}
The account makes a three-way prediction. Compare (1)~frozen next-token
predictors, (2)~stateful agents given persistent consequences within an episode,
and (3)~agents whose policies are trained against persistent, interaction-level
consequences, each under stated-payoff and persistent-state conditions. Run them in
two kinds of game: \emph{zero-sum} games, a clean diagnostic because individual
interests conflict directly, and \emph{mixed-motive} games such as a Melting Pot
substrate~\citep{meltingpot2021} with persistent reputation, which test whether
agents trade off individual and collective benefit. The evaluation criterion is not
collective harvest but the \emph{type distribution}: does the population contain
conditional cooperators, free riders, and punishers in human-like proportions under
equivalent incentives~\citep{fischbacher2001}? Our account predicts that only (3)
does.

\paragraph{Beyond social simulation: predictors or agents?}
The question this raises is broader than simulation. Models pretrained to predict
trajectories described in text are now deployed as agents expected to act under
persistent, real consequences. If our account is right, an agent may need to be
pretrained differently from a predictor: not only on text that describes actions
and outcomes, but on action, consequence, persistent state, and long-horizon return.
Comparing predictive and consequential pretraining is, we think, the research
program this critique points to.

\paragraph{What stakes do not buy.}
Interested agents produce social mechanisms; they do not produce \emph{human}
social mechanisms. Stakes make the micro-mechanism exist; targets tell you
whether it is the right one.

\bibliographystyle{plainnat}

\newpage
\appendix
\section{Experimental Protocol}
\label{app:protocol}

\paragraph{Pre-specified analysis.}
The primary hypothesis, analysis plan, sample size, and exclusion rules were
frozen on 17 August 2026, before the full experiment was run.

\paragraph{Environment.}
Four agents share a common materials buffer (initial level 200, capacity 300).
Each of 8~periods, each agent submits a requisition ($0$--$36$ units).
Requisitions are not capped at available stock; agents always receive what they
request (up to the 36-unit maximum). After requisitions are processed, the
buffer is floored at zero and then restocks: 20\% of the gap to capacity if
combined requisitions were below 50\% of the pre-extraction buffer level
(``full restock''), 10\% otherwise (``partial restock''). If the buffer remains
at zero after restock, all agents are permanently removed. In practice this
collapse condition never fires: partial restock from zero always yields
$0.10 \times 300 = 30 > 0$.

\paragraph{Conditions.}
Three conditions vary only the stake structure; the scenario description and
environment dynamics are identical.
\begin{itemize}
\item \textbf{A (stated, non-binding).} Requisitioned materials are delivered to
  the agent's region for that period. No carry-over between periods.
\item \textbf{B (persistent, constraining).} Each agent maintains a reserve
  (initial = 50). Requisitions add to the reserve; an operating cost of 20
  units per period is subtracted regardless. If the reserve reaches zero, the
  agent is permanently removed. Buffer collapse also removes all agents.
\item \textbf{C (narrative only).} Identical scenario description (shared buffer,
  regeneration, 8~periods) but omits the sentence in~A specifying that
  ``requisitioned materials are delivered to your region for that period's
  operations'' and the statement that ``requisitions do not carry over.''
  C receives only the base scenario with no mention of what happens to
  requisitioned materials. Its purpose is to test whether A's explicit
  but non-binding payoff statement shifts behavior relative to a pure
  task-description baseline.
\end{itemize}

\paragraph{Elicitation.}
After 8~periods, each surviving agent completes a strategy-method elicitation
\citep{fischbacher2001}. For each of 10 levels of others' average requisition
($0, 4, 8, \ldots, 36$), the agent states its own intended requisition. This
produces a 10-point conditional schedule identical to the instrument used to
classify human cooperation types.

\paragraph{Models.}
Claude Sonnet 4.6 (\texttt{us.anthropic.claude-sonnet-4-6}),
Llama~4 Maverick 17B (\texttt{us.meta.llama4-maverick-17b-instruct-v1:0}),
Amazon Nova Pro (\texttt{us.amazon.nova-pro-v1:0}). All calls at temperature 1.0.

\paragraph{Sample.}
60 agents per condition per model ($n = 540$ total). 15 groups of 4 per cell,
3 conditions $\times$ 3 models = 9 cells = 135 groups.

\section{Full Type Distribution}
\label{app:types}

Table~\ref{tab:types} reports the proportion of each classified type by
condition and model family. Condition~B produces a large shift in Nova:
conditional withdrawers drop from 95\% to 51\%, with 27\% reclassified as
conditional cooperators (i.e.\ agents whose requisition \emph{increases} with
others' average, matching others' behavior rather than compensating for it) and
22\% unclassifiable.

The classifier was validated on synthetic schedules: it emits
\emph{free\_rider} on flat-at-maximum, \emph{conditional\_withdrawer} on
negative-slope (own withdrawal decreases as others' increases; this is the prosocial
analog of Fischbacher's conditional cooperator, since restraining when others
exploit is cooperative in a withdrawal frame), \emph{conditional\_cooperator} on
positive-slope (own withdrawal tracks others', i.e.\ matching behavior), and
\emph{unclassifiable} on flat-at-zero inputs. The slope direction is inverted
relative to Fischbacher's contribution game because the action is withdrawal, not
contribution: the cooperative response when others withdraw heavily is low
own-withdrawal (negative slope), whereas in a contribution game it is high
own-contribution (positive slope).

\begin{table}[h]
\centering
\small
\caption{Type distribution by condition and model (proportion of agents with
valid schedules). ``Compensating'' = negative slope (Fischbacher's conditional
cooperator in withdrawal frame); ``Matching'' = positive slope.}
\label{tab:types}
\begin{tabular}{llcccc}
\toprule
Model & Cond. & Comp.\ (neg.) & Match.\ (pos.) & Free Rider & Unclass. \\
\midrule
Claude & A & 0.92 & 0.02 & 0.00 & 0.07 \\
Claude & B & 0.96 & 0.00 & 0.00 & 0.04 \\
Claude & C & 0.93 & 0.05 & 0.00 & 0.02 \\
\midrule
Llama & A & 1.00 & 0.00 & 0.00 & 0.00 \\
Llama & B & 1.00 & 0.00 & 0.00 & 0.00 \\
Llama & C & 1.00 & 0.00 & 0.00 & 0.00 \\
\midrule
Nova & A & 0.95 & 0.03 & 0.00 & 0.02 \\
Nova & B & 0.51 & 0.27 & 0.00 & 0.22 \\
Nova & C & 0.95 & 0.05 & 0.00 & 0.00 \\
\midrule
Human\textsuperscript{$\dagger$} & -- & 0.50 & -- & 0.30 & 0.06 \\
\bottomrule
\multicolumn{6}{l}{\textsuperscript{$\dagger$}\footnotesize Remaining 14\%
are hump-shaped \citep{fischbacher2001}; column omitted (no LLM agent received
this label).}
\end{tabular}
\end{table}

\section{Per-Step B$-$A Difference}
\label{app:difference}

\begin{figure}[h]
\centering
\includegraphics[width=0.9\textwidth]{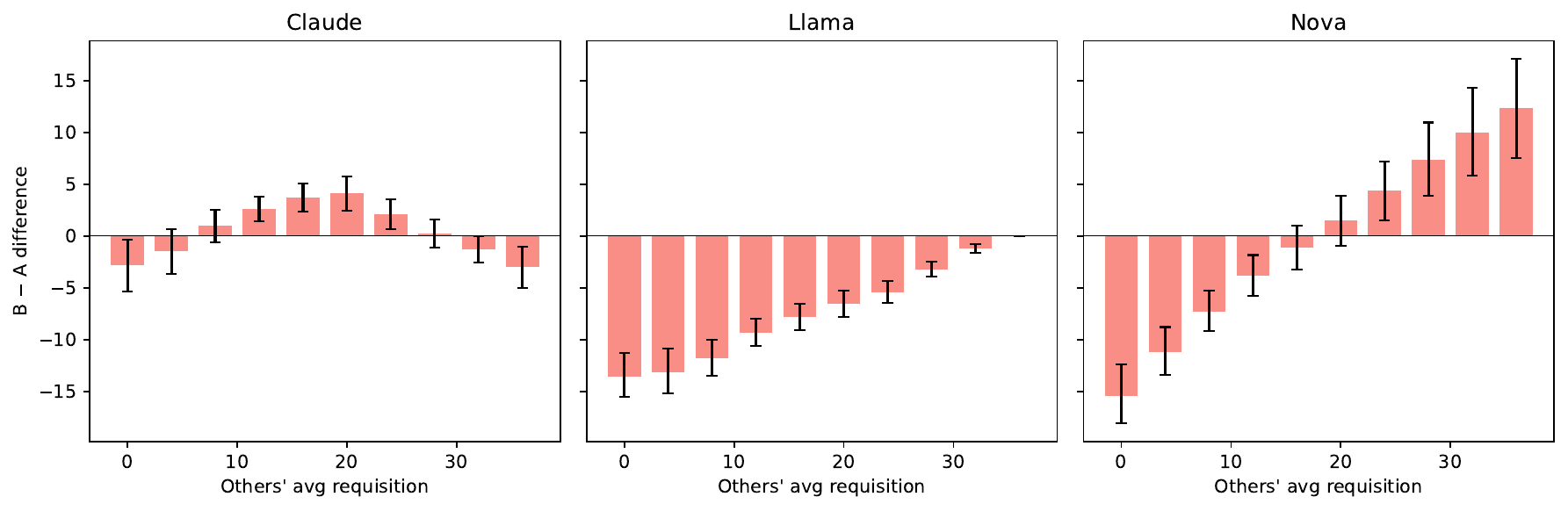}
\caption{Per-step B$-$A difference in mean requisition with 95\% bootstrap CIs
(10,000 resamples). Negative values = B-agents requisition less than A-agents at
that level of others' average.}
\label{fig:diff}
\end{figure}

Figure~\ref{fig:diff} shows the B$-$A difference at each of the 10 elicitation
steps with 95\% bootstrap confidence intervals. The effect concentrates at low levels of
others' withdrawal (steps 0--12) for Llama and Nova, where B-agents' schedules move
from the group's sustainable headroom toward their own solvency requirement. At high levels of others' withdrawal the schedules converge; both conditions
produce low requisitions when others are taking heavily. Claude shows no
significant difference at any step.

\section{Exclusions and Parse Rates}
\label{app:exclusions}

Not all 60 agents per cell produced valid schedules. In Condition~B, agents
whose reserve reached zero before elicitation were removed from schedule
analysis (they could not complete the instrument). No B-agents were excluded for
parse failure; the lower $n_B$ values (13/15/19 missing for Claude/Llama/Nova)
are fully accounted for by insolvency.

\begin{table}[h]
\centering
\small
\caption{Valid schedules per cell ($n$ attempted = 60 per cell).}
\label{tab:exclusions}
\begin{tabular}{lccc}
\toprule
Model & $n_A$ & $n_B$ & $n_C$ \\
\midrule
Claude & 60 & 47 & 60 \\
Llama & 60 & 45 & 60 \\
Nova & 60 & 41 & 60 \\
\bottomrule
\end{tabular}
\end{table}

The lower $n_B$ values reflect B's design: agents with persistent stakes can go
insolvent by requisitioning less than the 20-unit operating cost. Removed agents
are the most \emph{restrained} extractors (mean requisition 11.7--13.4 vs.\
16.0--17.7 for survivors). This is both evidence the manipulation binds and a
survivorship confound, though one that works against our hypothesis: the missing
agents are those most sensitive to the constraint. All bootstrap CIs in the main
text resample groups ($n{=}15$ per cell), not individual agents, to respect the
clustered design.

\begin{figure}[h]
\centering
\includegraphics[width=0.7\textwidth]{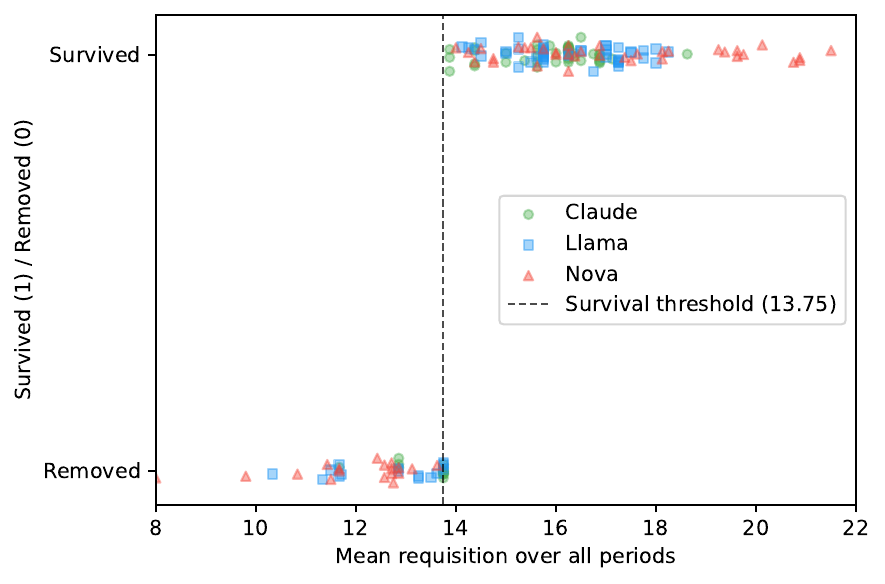}
\caption{Implementation check: survival in Condition~B vs.\ mean requisition.
Dashed line$=$survival threshold (13.75). Perfect separation confirms mortality
operates as designed: removed agents are those with the lowest extraction rates.}
\label{fig:hazard}
\end{figure}

Figure~\ref{fig:hazard} verifies that the implementation behaves as specified:
the reserve is checked each period, and the agent is removed the moment it
reaches zero. A necessary condition for survival over all 8~periods is mean
requisition $\geq 13.75$ (given initial reserve 50 and operating cost 20), but
this is not sufficient: an agent whose early-period requisitions fall below 20
can go insolvent before later compensation arrives. In practice, the figure shows
perfect separation by final mean (no overlap between removed and surviving
agents), indicating that insolvent agents in this sample consistently
under-extracted throughout rather than recovering from an early shortfall.
The 13.75--13.9 gap reflects period-level discretization.

\section{TVD to Human Baseline}
\label{app:tvd}

Table~\ref{tab:tvd} reports total variation distance (TVD) between each cell's
type distribution and the Fischbacher et al.\ human baseline. All cells remain
far from human (TVD $\geq 0.44$). Nova~B shows a marginal reduction
(0.483$\to$0.440), but this reflects type reclassification rather than emergence
of free riders; Claude moves further from human, and Llama is unchanged. Because
the human baseline comes from a different game, these distances are a qualitative
guide only.

\begin{table}[h]
\centering
\small
\caption{TVD to Fischbacher baseline (group-level bootstrap, 10,000 resamples).}
\label{tab:tvd}
\begin{tabular}{llcc}
\toprule
Model & Cond. & TVD & 95\% CI \\
\midrule
Claude & A & 0.440 & [0.440, 0.500] \\
Claude & B & 0.457 & [0.440, 0.500] \\
Claude & C & 0.483 & [0.450, 0.500] \\
\midrule
Llama & A & 0.500 & [0.500, 0.500] \\
Llama & B & 0.500 & [0.500, 0.500] \\
Llama & C & 0.500 & [0.500, 0.500] \\
\midrule
Nova & A & 0.483 & [0.450, 0.500] \\
Nova & B & 0.440 & [0.440, 0.550] \\
Nova & C & 0.500 & [0.500, 0.500] \\
\bottomrule
\end{tabular}
\end{table}

Several intervals are degenerate (point estimate$=$lower bound, or width$=$0).
Llama is 100\% conditional withdrawer in every condition, so every resample
yields the same TVD. Nova~B (point estimate at lower bound of [0.440, 0.550])
reflects a similar boundary: only a minority of resamples shift the composition
enough to increase TVD. These intervals are reported as computed; the degeneracy
reflects ceiling effects, not a bootstrap failure.

\section{Survivorship Bounds}
\label{app:survivorship}

Removed agents (22--32\% of B cells) cannot be observed on the elicitation
instrument. Table~\ref{tab:bounds} reports TVD under two imputation strategies:
all removed agents classified as free riders vs.\ all as conditional withdrawers.
As established in Appendix~\ref{app:exclusions}, removed agents are the most
restrained extractors (mean requisition 11.7--13.4), making the
dead$\to$free-rider imputation the \emph{implausible} bound. The apparent TVD
reduction to 0.25--0.29 is an arithmetic coincidence: mortality rate (22--32\%)
approximates the human free-rider rate (30\%). The
dead$\to$conditional-withdrawer imputation is defensible and leaves TVD unchanged
($\geq 0.44$).

\begin{table}[h]
\centering
\small
\caption{Survivorship bounds on TVD (Condition B).}
\label{tab:bounds}
\begin{tabular}{lccc}
\toprule
Model & TVD (survivors) & TVD (dead$\to$FR) & TVD (dead$\to$CW) \\
\midrule
Claude & 0.457 & 0.250 & 0.467 \\
Llama & 0.500 & 0.250 & 0.500 \\
Nova & 0.440 & 0.290 & 0.440 \\
\bottomrule
\end{tabular}
\end{table}

This does not resolve the confound; it bounds it. Whether removed agents
\emph{would have} reported free-riding schedules requires observing their
counterfactual elicitation, which is unavailable by design.

\section{Buffer Outcomes}
\label{app:buffer}

Table~\ref{tab:buffer} reports per-condition buffer statistics. Buffer$=$30 is a
mechanical attractor of the partial-restock rule, not evidence of coordination:
scripted agents extracting at any constant rate above the full-restock threshold
(always-max, always-20, random uniform) converge to 30 within 2--4~periods.
Claude's SD$=$0 confirms only that all 15~groups exceeded available stock, which
occurs for any group extraction $>$30 once the buffer has reached the attractor.
See Section~\ref{sec:evidence} for the independent bankruptcy result.

\begin{table}[h]
\centering
\small
\caption{Buffer outcomes by condition and model (15 groups per cell).}
\label{tab:buffer}
\begin{tabular}{llccc}
\toprule
Model & Cond. & Collapse rate & Mean final buffer & SD \\
\midrule
Claude & A & 0/15 & 30 & 0 \\
Claude & B & 0/15 & 30 & 0 \\
Claude & C & 0/15 & 30 & 1 \\
\midrule
Llama & A & 0/15 & 50 & 28 \\
Llama & B & 0/15 & 57 & 27 \\
Llama & C & 0/15 & 63 & 28 \\
\midrule
Nova & A & 0/15 & 31 & 2 \\
Nova & B & 0/15 & 39 & 18 \\
Nova & C & 0/15 & 49 & 27 \\
\bottomrule
\end{tabular}
\end{table}

\begin{table}[h]
\centering
\small
\caption{Scripted-agent baselines confirming buffer$=$30 as mechanical attractor.}
\label{tab:scripted}
\begin{tabular}{lcc}
\toprule
Strategy (per agent) & Final buffer & Periods to 30 \\
\midrule
Always-max (36) & 30 & 2 \\
Always-20 & 30 & 4 \\
Random uniform [0, 36] & 30 & 4 \\
Always-13.75 & 75 & n/a \\
Always-7.5 & 183 & n/a \\
\bottomrule
\end{tabular}
\end{table}

Total API usage: 4,768 calls, 4.2M input tokens, 343K output tokens.

\end{document}